\documentclass[conference]{IEEEtran}
\usepackage[T1]{fontenc}
\usepackage[utf8]{inputenc}
\usepackage{lmodern}
\usepackage{amsmath}
\usepackage{booktabs}
\usepackage{graphicx}
\usepackage{tikz}
\usetikzlibrary{arrows.meta,positioning,fit}
\usepackage{url}
\usepackage[hidelinks]{hyperref}

\begin{document}

\title{Silinx: An Open, Browser-Based\\HDL Design Platform for Teaching Digital Systems\\on FPGA Boards}

\author{\IEEEauthorblockN{Pedro Mal\'o}
\IEEEauthorblockA{Departamento de Engenharia Electrot\'ecnica e de Computadores (DEEC)\\
Faculdade de Ci\^encias e Tecnologia (FCT)\\
Universidade NOVA de Lisboa, Portugal}}

\maketitle

\begin{abstract}
Many digital systems courses teach on FPGA boards whose vendor tool is no longer maintained: the Digilent
Basys2 and Nexys2 (Xilinx Spartan-3E) and Nexys3 and Atlys (Spartan-6) need Xilinx ISE~14.7, discontinued
since 2013 and hard to install on the computers and tablets students use today. The author presents Silinx, a free
(AGPL-3.0) HDL design platform for teaching that keeps ISE's Project Navigator workflow but runs in a web
browser with nothing to install. A module can be written in VHDL or Verilog, in C or C++ through Silinx
HLS, which is compatible with AMD's Vitis HLS and passes all 73 of its introductory examples, or drawn as
a schematic, an ASM chart, a state diagram, a Boolean function with its Karnaugh map, or a Petri net,
from which its HDL is generated. Designs are checked with explanations written for beginners, simulated,
run on an emulated board, built, and programmed onto a real board over WebUSB. The implementation flow is
open and runs in the browser: Yosys and nextpnr compiled to WebAssembly, with Silinx's own device data,
timing model and bitstream writer for the Spartan-3E and Spartan-6, built by black-box observation of
ISE's documented tools; Lattice iCE40 boards use the existing open iCE40 tools, and further families are
planned. ISE is the oracle: Silinx's bitstreams are byte-identical to those ISE writes for the
same placed and routed design, and pass ISE's design rule check. Against ISE at the same optimization goal, the open flow uses 15\% fewer slices with 14\% shorter critical
paths when optimizing for area, and 21\% fewer slices with 8\% shorter critical paths when optimizing for
speed (geometric means over 45 open-source designs built by both flows).
The author also describes how the platform was developed with AI coding agents under
test-first project rules.
\end{abstract}


\section{Introduction}

Introductory digital systems courses teach combinational logic, state machines and their description in
a hardware description language (HDL), and end with a circuit running on a real FPGA board. Many such
courses use Digilent's Basys2 and Nexys2 boards, built around the Xilinx Spartan-3E, or the Nexys3 and
Atlys, built around the Spartan-6. The boards still work and teach the subject well, and replacing them
for a whole class is expensive. Their software has not kept up: both families are supported only by
Xilinx ISE~14.7, whose development stopped in 2013, and not by its successor, Vivado. ISE does not run
on macOS or natively on Apple silicon, installs awkwardly on recent Windows and Linux, is a heavy
download for a student's laptop, and nothing in it will be fixed. Course time that should go into
digital design goes into installing the tools.

Silinx is the author's answer to this problem. It is an HDL design platform, and its toolchain is one part of it: projects of
VHDL, Verilog and C/C++ sources and of drawings, editors with help for beginners, a simulator, a board
emulator, implementation and board programming. It follows the ISE~14.x Project Navigator, so that
course material, lab guides and teachers' habits still apply, but it runs in a web browser, including
synthesis, placement, routing, bitstream generation and programming. Its open toolchain builds for the
Basys2 (XC3S100E, XC3S250E), Nexys2 (XC3S500E, XC3S1200E), Nexys3 (XC6SLX16) and Atlys (XC6SLX45),
and for the Lattice iCE40 of the iCEBreaker; other boards can use a local ISE~14.7 through Silinx's
local server, and further families are planned. A student opens a web page and can go from a drawing to
a working board without installing anything.

The contributions of this paper are:
\begin{itemize}
\item a browser-based HDL design platform for an introductory course, in which drawings (schematics,
ASM charts, state diagrams, Boolean functions, Petri nets) and C/C++ functions are the base of a module
and its HDL is generated from them (Section~\ref{sec:env});
\item an open implementation flow for the Spartan-3E and Spartan-6 boards, from HDL to bitstream with no
Xilinx tool at any step, built on Yosys and nextpnr and on a bit database obtained by black-box
observation of ISE's documented tools (Section~\ref{sec:toolchain});
\item a verification method that uses ISE as an oracle: byte-identical bitstreams, a clean design rule
check (DRC), and the course's examples as the regression suite (Section~\ref{sec:verif});
\item a quality-of-results comparison with ISE on an open-source design corpus, per optimization goal
(Section~\ref{sec:eval}); and
\item an account of how the platform was developed with AI coding agents under explicit, test-first
project rules (Section~\ref{sec:ai}).
\end{itemize}

\section{Background and Related Work}

\paragraph*{Educational design tools}
Logisim and its successors, Digital and CircuitVerse, are widely used to teach logic with simulated
schematics; where they reach an FPGA board, they rely on the vendor's tools to do so. Online simulators
such as EDA Playground run HDL simulations in the browser but do not target a board. Icestudio~\cite{icestudio}
provides graphical design for Lattice iCE40 boards on top of the open IceStorm flow. None of these keeps
the workflow of ISE, which existing Spartan-3E and Spartan-6 courses are written for, and none produces
bitstreams for these families.

\paragraph*{Open FPGA toolchains}
Project IceStorm~\cite{icestorm} documented the Lattice iCE40 bitstream, Project Trellis~\cite{trellis}
the Lattice ECP5 and Project X-Ray~\cite{xray} the Xilinx 7-series, each by building many small designs
with the vendor's tools and comparing the resulting bitstreams. With Yosys~\cite{yosys} for synthesis and
nextpnr~\cite{nextpnr} or VPR~\cite{vtr} for place and route, they form complete open flows; YoWASP~\cite{yowasp}
compiles several of these tools to WebAssembly. Earlier work recovered parts of the bitstream formats of
older Xilinx families to recover netlists from bitstreams~\cite{debit,bil}. XDL, the text design format
ISE provides for users' own tools, was the basis of research tool kits such as Torc~\cite{torc} and
RapidSmith~\cite{rapidsmith}. To the author's knowledge no open flow produced Spartan-3E bitstreams before
Silinx.

\section{The Learning Environment}
\label{sec:env}

\begin{figure*}[t]
\centering
\includegraphics[height=56mm]{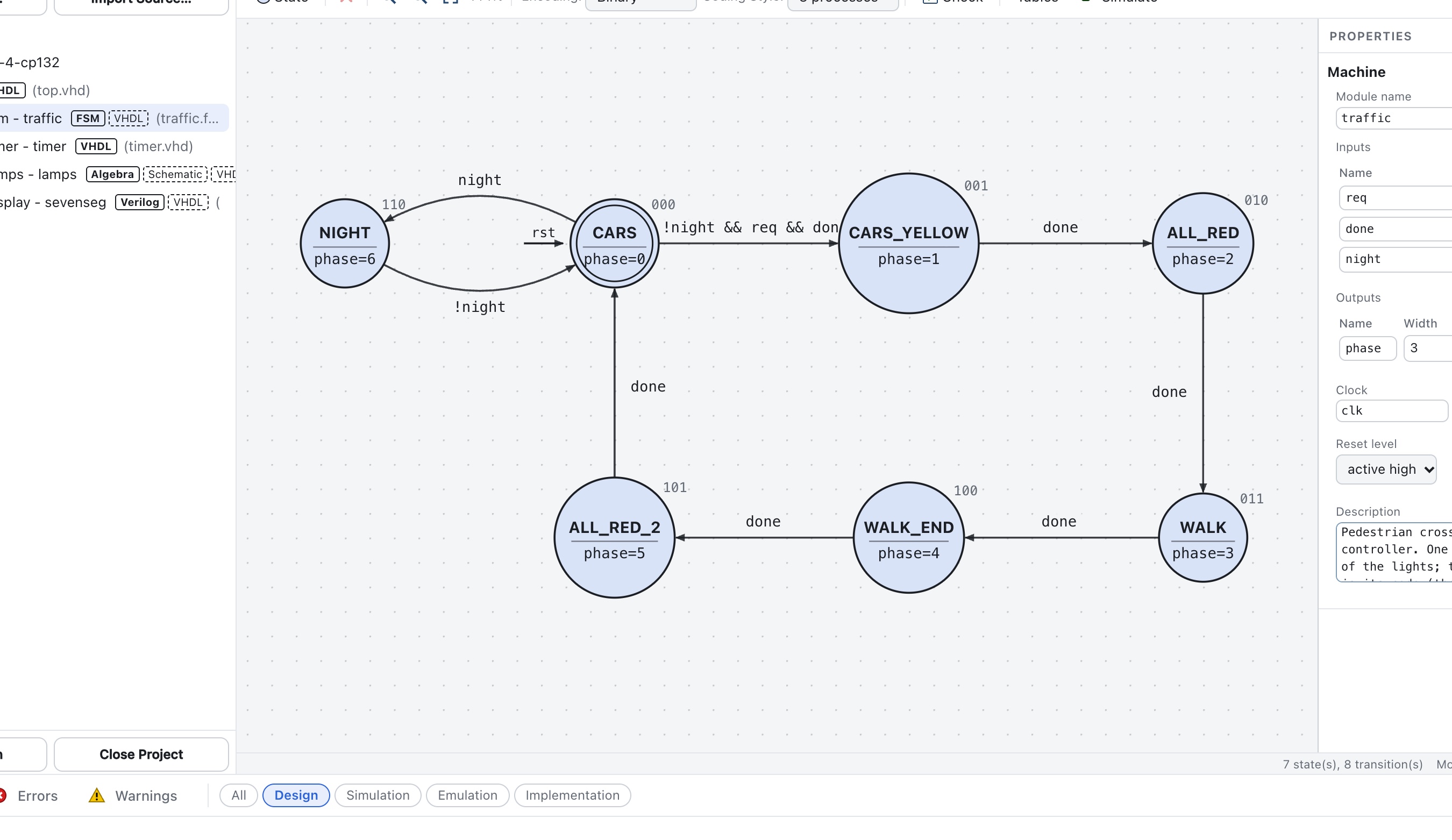}\hfill
\includegraphics[height=56mm]{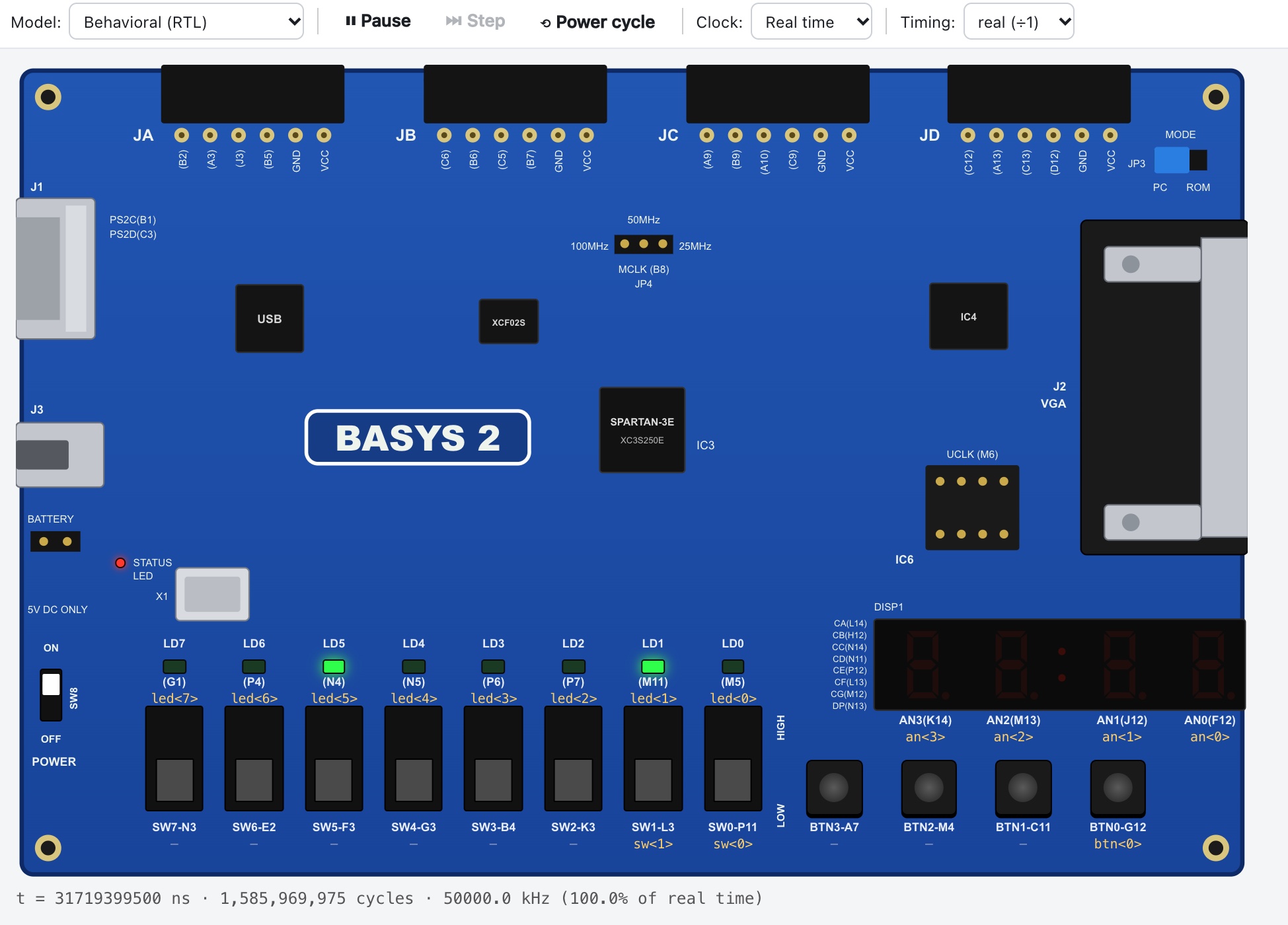}\\[0.5mm]
\makebox[0.56\textwidth]{\footnotesize (a) State diagram editor}\hfill
\makebox[0.43\textwidth]{\footnotesize (b) Board Emulator}
\caption{Silinx in the browser, on the \emph{traffic} example (a pedestrian crossing on a Basys2).
(a)~The controller drawn as a state diagram, from which its HDL is generated; the Design pane lists one
row per module, with badges for its drawing first and its generated HDL last. (b)~The whole design on a
drawing of the Basys2, wired by its constraints file (switches, buttons, LEDs, seven-segment display), on
the compiled engine. Screenshots of the public site, release 15.23.0.}
\label{fig:ui}
\end{figure*}

\paragraph*{An ISE-like project navigator, anywhere}
Projects are made of VHDL and Verilog modules (mixed language), with implementation and simulation
roles, a UCF constraints file, and import and export of ISE projects (\texttt{.xise}, including ISE
schematics). The processes (Check Syntax, Synthesize, Implement Design, Generate Programming File,
Configure Target Device), the design summary and the reports follow ISE, so that lab guides written for
ISE still apply. The same page has two interfaces, a modern one organised around a lab
session's tasks and a classic, ISE-like one, in light and dark themes, in English and Portuguese, with
touch and narrow screens supported. The same application runs as a public site (installable as an app,
also offline), as a single HTML file, or as a local Node.js server, on Apple-silicon Macs, Windows,
Linux, Chromebooks and Android tablets.

\paragraph*{Drawings are the base, HDL is generated}
First-semester courses start from drawings and come to code later. Silinx has a schematic editor in the style of
ISE's (with ISE \texttt{.sch} import and export and a live, Logisim-style simulation of the drawing), an
ASM chart editor with step-by-step simulation, a state diagram (Moore/Mealy) editor, a Boolean algebra
editor, and a Petri net editor. The Boolean algebra editor takes a truth table or an expression per
output and shows its Karnaugh map with the groups of the minimal sum of products or product of sums
(Quine--McCluskey with exact cover); when a function has several minimal solutions, each can be browsed
beside the map and the one chosen is used in the HDL and the schematic. The canonical forms are shown in
full with their cost, and maps of three and four variables can be shown in 3D, on a cylinder and on a
torus, where the adjacency that wraps around the flat map's edges becomes visible. Each drawing
always has its VHDL or Verilog module, generated from it, so that the drawing stays the single
source; the Design pane shows one row per module, its drawing first (Fig.~\ref{fig:ui}a).
Conversely, an HDL module can be converted to a schematic, and an HDL state machine to a state diagram or
an ASM chart. A student thus sees, at every step, the HDL that corresponds to what they drew.

\paragraph*{C and C++ as sources}
A C or C++ function can be a design source through \emph{Silinx HLS}, which follows AMD's Vitis HLS
(which does not target these families): the generated VHDL or Verilog module has the ports and control
protocol Vitis HLS would give it, and the C is the module's base, as a drawing is. It supports the
\texttt{PIPELINE}, \texttt{UNROLL}, \texttt{ARRAY\_PARTITION}, \texttt{ARRAY\_RESHAPE},
\texttt{DATAFLOW}, \texttt{ALLOCATION} and \texttt{EXPRESSION\_BALANCE} pragmas, \texttt{hls::stream}
FIFOs and RTL blackboxes. The test bench's \texttt{main()} runs as a C simulation, and its calls are
replayed on the generated HDL as a C/RTL co-simulation. On AMD's 73 Vitis HLS introductory examples, all
73 pass C simulation, synthesis and co-simulation. Students can thus meet high-level synthesis on the
boards they already have.

\paragraph*{Help for beginners}
The common errors (a missing \texttt{;} or \texttt{end}, an undeclared signal, \texttt{std\_logic}
vs.\ \texttt{std\_logic\_vector}, width mismatches, a missing \texttt{use} clause, several drivers)
keep their ISE-style message but come with an \emph{Explanation} and a \emph{How to fix}, in English or
Portuguese, and many have quick fixes in the editor. Check Syntax and the live editor also run the
design checks an instructor would otherwise point out by hand: an inferred latch, an incomplete
sensitivity list, a combinational loop, an output never assigned, a clock used as data, and blocking and
non-blocking assignments in the wrong kind of Verilog \texttt{always} block. The editor also has
symbol renaming and go-to-definition.

\paragraph*{Simulation and the Board Emulator}
SSim, Silinx's own simulator, is event-driven and four-valued, with delta cycles, VHDL and Verilog test
benches, an ISim-like waveform window, stimuli drawn on the waveforms (as ISE's HDL Bencher did), and a
wizard that writes self-checking test benches. The \emph{Board Emulator} runs the design on a drawing of
the Basys2, Nexys2, Nexys3, Atlys, Spartan-3E Starter Kit or iCEBreaker, wired by the constraints:
students click switches and buttons and watch LEDs and the multiplexed seven-segment display
(Fig.~\ref{fig:ui}b); a VGA monitor shows the picture a design scans out. Besides an event-driven engine,
a compiled engine reads the design at word level through Yosys and runs it as one JavaScript function per
clock cycle. On the course examples it runs at 59 to 141 million cycles per second (20 for an 8-bit
microprocessor), measured in Node.js on an Apple M4 Pro, so counters, clock dividers and debouncers run
at the boards' 50\,MHz with their real values. The netlists after synthesis, mapping and place and route
can be emulated and simulated too, the routed one in SSim with its delays from an SDF file, so that a
clock too fast for the routed design shows up in simulation.

\paragraph*{Programming the board from the browser}
The Digilent Adept USB protocol and the JTAG programming of the FPGA and of the board's Platform Flash
are implemented in JavaScript over WebUSB, so Chrome or Edge on a computer, or Chrome on an Android
tablet, programs the board directly; the iCEBreaker's SPI flash is programmed the same way. Finally, a
Model Context Protocol (MCP) server lets an AI assistant perform the same operations as the interface on
a student's or teacher's projects.

\section{The Open Toolchain}
\label{sec:toolchain}

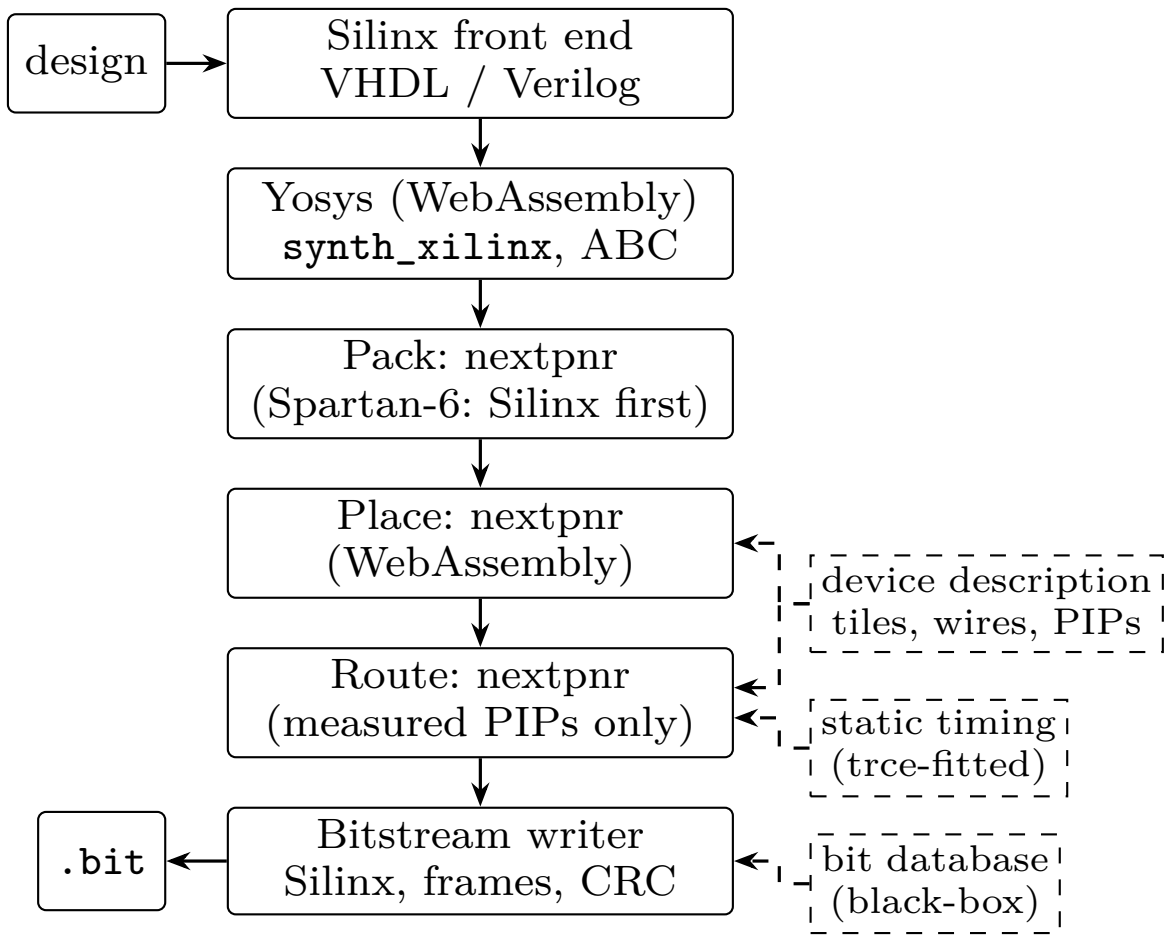
\begin{figure}[t]
\centering
\begin{tikzpicture}[
  node distance=3.2mm,
  box/.style={draw, rounded corners=1.5pt, align=center, font=\footnotesize,
              text width=32mm, minimum height=6.5mm, inner sep=2pt},
  data/.style={draw, dashed, align=center, font=\scriptsize, inner sep=2pt},
  arr/.style={-{Stealth[length=1.6mm]}, semithick}]
\node[box] (fe) {Silinx front end\\VHDL / Verilog};
\node[box, below=of fe] (ys) {Yosys (WebAssembly)\\\texttt{synth\_xilinx}, ABC};
\node[box, below=of ys] (pk) {Pack: nextpnr\\(Spartan-6: Silinx first)};
\node[box, below=of pk] (pl) {Place: nextpnr\\(WebAssembly)};
\node[box, below=of pl] (rt) {Route: nextpnr\\(measured PIPs only)};
\node[box, below=of rt] (bg) {Bitstream writer\\Silinx, frames, CRC};
\draw[arr] (fe) -- (ys); \draw[arr] (ys) -- (pk); \draw[arr] (pk) -- (pl);
\draw[arr] (pl) -- (rt); \draw[arr] (rt) -- (bg);
\node[data, right=5mm of pl, yshift=-4mm] (dev) {device description\\tiles, wires, PIPs};
\node[data, right=5mm of bg, yshift=-1.5mm] (db) {bit database\\(black-box)};
\node[data, right=5mm of rt, yshift=-3mm] (sta) {static timing\\(trce-fitted)};
\draw[arr, dashed] (dev.west) -- ++(-2mm,0) |- (pl.east); \draw[arr, dashed] (dev.west) -- ++(-2mm,0) |- ([yshift=1mm]rt.east);
\draw[arr, dashed] (db.west) -- ++(-2mm,0) |- (bg.east); \draw[arr, dashed] (sta.west) -- ++(-2mm,0) |- ([yshift=-1mm]rt.east);
\node[box, left=4mm of bg, text width=7mm] (bit) {\texttt{.bit}};
\draw[arr] (bg) -- (bit);
\node[box, left=4mm of fe, text width=9mm] (src) {design};
\draw[arr] (src) -- (fe);
\end{tikzpicture}
\caption{The open implementation flow for the Spartan-3E and Spartan-6. Every stage runs in the browser
(in Web Workers) or in Node.js; no Xilinx tool is used. The iCE40 uses Yosys, nextpnr-ice40 and
icepack.}
\label{fig:flow}
\end{figure}

Fig.~\ref{fig:flow} shows the flow. Yosys~\cite{yosys} runs as WebAssembly (YoWASP~\cite{yowasp}), and
nextpnr~\cite{nextpnr} as WebAssembly builds of a fork of its Himbaechel architecture framework~\cite{nextpnrspartan}
in which Silinx's Spartan-3E and Spartan-6 micro-architectures are written; the rest is JavaScript, so
the same code runs in the browser and on a server. The toolchain is also published on npm as a package of
its own (\texttt{silinx-fpga}, Apache-2.0).

\subsection{Front end and synthesis}
Silinx's own VHDL and Verilog parsers and elaborator, already needed by the editor and the simulator,
write the elaborated design as one flat Verilog module, so the toolchain never reads VHDL itself; Yosys
synthesizes it with \texttt{synth\_xilinx -family xc3se} and ABC. Three optimization goals, chosen in the
process properties as in ISE, decide how hard the flow works:
\emph{Area}, the default of a new project, keeps the smaller of the usual mapping's netlist and the
mappings for speed, and places once; \emph{Speed} uses the mappings for speed and makes several
placements (32 for a design of up to 100 slices, 8 up to 500, 4 above), each routed, and keeps the fastest
by Silinx's timing analysis; \emph{Fast} uses the usual mapping and one placement, except that a deep
design (more than 16 logic levels, at most 1\,500 LUTs) also gets Speed's synthesis, kept if it is 10\%
faster.

\subsection{The bit database}
\label{sec:bitdb}
The bitstream writer needs to know which configuration bits set each feature of the chip: each LUT
memory bit, each multiplexer setting inside a slice, each I/O setting and each routing switch (PIP).
Following IceStorm, Trellis and X-Ray, Silinx builds this database by observing the vendor's tools.
Small test designs that differ in a single feature are written in XDL, ISE's documented text design
format; ISE's command-line tools turn them into bitstreams (\texttt{xdl -xdl2ncd}, then \texttt{bitgen}),
and the bits that change between two bitstreams are those of the feature that changed. The packets,
registers and frames follow Xilinx's public documentation (UG332, XAPP452); for the XC3S250E, the
configuration data is one write of 578 frames of 73 32-bit words. The other Spartan-3E devices share
the tile types' bits with the XC3S250E; the Spartan-6 database was measured the same way. Every
experiment and analysis script is kept in the repository, so that anyone with a lawful copy of ISE can
rebuild and check the database.

No Xilinx program is decompiled or disassembled, no internal data file of ISE is read, no Xilinx file is
shipped, and no protection (bitstream encryption, readback protection, licence checks) is studied. The
method relies on the right of a lawful user to observe, study and test how a program functions (EU
Directive 2009/24/EC, Art.~5(3)); which bit configures which feature is a fact about the hardware's
interface rather than an expression of ISE's code.

\subsection{Device description and timing}
Placement and routing need the chip's tiles, sites, wires and switches. Silinx computes a compact
description from the output of ISE's documented \texttt{xdl -report -pips -all\_conns} command. For the
XC3S250E the report is 204\,MB of text: 1\,419 tiles, 4\,615 sites, 569\,094 tile wires and
2\,139\,368 PIPs. Since most tiles of a kind are identical, the description stores each distinct tile
content once (293 templates; the 576 CLB tiles share one) and the connections between tiles as 200\,181
nodes; gzipped, it is 1.96\,MB. The descriptions of the six devices of the supported boards are shipped,
so the flow needs no ISE; the four Spartan-3E dies have also been measured by black-box observation, and
designs routed on the measured descriptions pass ISE's DRC with byte-identical bitstreams. nextpnr's
chip database is generated from these descriptions, restricted to the PIPs whose bits are known, with
Silinx's delay model, which is fitted to ISE's timing analyser (\texttt{trce}).

\subsection{Packing, placement and routing}
nextpnr packs, places and routes, timing-driven, at every goal; on the Spartan-6, Silinx's own packer
runs first. The routed design is written back as XDL, so the bitstream writer and the ISE oracle of
Section~\ref{sec:verif} see the same format as before. Until late October 2026, placement and routing
were done by Silinx's own simulated-annealing placer~\cite{tdplace} and PathFinder-style
router~\cite{pathfinder}; nextpnr became the default in release 15.21.0, and the only engine in 15.22.0,
once it had reached parity, while the Spartan-6 packer and the bitstream writer remain Silinx's.

\section{Verification: ISE as the Oracle}
\label{sec:verif}

An open flow for a closed device is only useful if its bitstreams are right: a wrong bit can short two
drivers on a student's board. Silinx therefore treats ISE as the reference at every level.

\paragraph*{Byte-identical bitstreams}
The design placed and routed by Silinx is written out as XDL and given to ISE, which converts it
(\texttt{xdl -xdl2ncd}), runs its DRC and writes its own bitstream with \texttt{bitgen}. Silinx's
bitstream for the same design must be \emph{byte-identical} to ISE's, and ISE's DRC must find no error
(\texttt{npm run check:open-flow}). This holds for the designs checked on the Spartan-3E and on the
Spartan-6 boards. Negative controls show that the check is meaningful: a design with all its routing
removed gives 372 DRC errors, and a single PIP leading nowhere gives an ``antenna'' error.

\paragraph*{The examples are the test designs}
Ten example projects form the course's progression and the regression suite: \emph{gates} (a
schematic), \emph{traffic} (a state diagram, a Boolean function, VHDL and Verilog), \emph{blinky} (an
ASM chart), \emph{pong}, \emph{stopwatch}, \emph{clock}, \emph{vga}, \emph{isqrt} (a module written in C,
made into hardware by Silinx HLS), \emph{uart} and \emph{micro80} (a microprocessor running a program
from block RAM). Each must build and run on every device of the open toolchain and build with ISE with no
warnings, and each has a self-checking test bench.

\paragraph*{Tests and continuous integration}
Every change ships with its own tests, and tests assert the exact expected warnings of each flow, never
only the absence of errors. A pull request runs the unit tests on Linux and the UI tests in both
interfaces and in Firefox and WebKit; the ISE tests run nightly and the comparison with ISE weekly, on a
self-hosted runner with a private ISE installation that is never published.

\section{Evaluation}
\label{sec:eval}

\paragraph*{Setup}
The design corpus collects 67 open-source designs: CPUs (T80, light8080, PicoRV32, NEORV32, SERV,
FemtoRV32), peripherals (UART, SPI, I2C, FIFOs), cryptographic cores (AES, SHA-1, SHA-256, SipHash),
floating-point units and benchmarks (EPFL~\cite{epfl}, ITC'99, IWLS'05, VTR~\cite{vtr}). Each design is
built by both flows for the same chip, the XC3S1200E-FG320 (Nexys2-1200), with the same top, at each of
Silinx's three optimization goals, and by ISE at the matching XST setting: Area against
\texttt{-opt\_mode Area -opt\_level~2}, Speed against \texttt{-opt\_mode Speed -opt\_level~2}, and Fast
against \texttt{-opt\_mode Speed -opt\_level~1}, XST's default. ISE's map and par are given their best
chance at every goal: map packs slices as tightly as it can (\texttt{-c 1}) and the clock constraint is
set to 5\,ns, so that its place and route works for speed. Both critical paths are measured by ISE's
\texttt{trce} on the routed design. The results were measured on 8 November 2026 (Silinx 15.22.1, ISE
14.7); the full tables are in the repository (\texttt{docs/QUALITY.md}).

\begin{table}[t]
\centering
\caption{Silinx vs.\ ISE per optimization goal on the XC3S1200E: geometric means of the ratios
Silinx/ISE over the 45 designs built by both flows (critical path: over those timed by both).
Negative: Silinx smaller, faster or quicker.}
\label{tab:qor}
\footnotesize
\setlength{\tabcolsep}{4pt}
\begin{tabular}{@{}lrrr@{}}
\toprule
 & Area & Speed & Fast \\
ISE (XST) setting & Area, 2 & Speed, 2 & Speed, 1 \\
\midrule
Logic LUTs              & $-16\%$ & $-21\%$ & $-19\%$ \\
Flip-flops              & $+5\%$  & $-9\%$  & $-4\%$  \\
Slices                  & $-15\%$ & $-21\%$ & $-20\%$ \\
As few LUTs or fewer    & 30/45   & 33/45   & 26/45   \\
\midrule
Critical path           & $-14\%$ & $-8\%$  & $+12\%$ \\
As fast or faster       & 29/39   & 28/40   & 10/39   \\
\midrule
Run time (indicative)   & $-80\%$ & $-11\%$ & $-84\%$ \\
\bottomrule
\end{tabular}
\end{table}

\begin{table}[t]
\centering
\caption{Selected designs: ratios Silinx/ISE of logic LUTs (L) and critical path (CP) at each goal.}
\label{tab:designs}
\footnotesize
\setlength{\tabcolsep}{3.2pt}
\begin{tabular}{@{}lrrrrrr@{}}
\toprule
 & \multicolumn{2}{c}{Area} & \multicolumn{2}{c}{Speed} & \multicolumn{2}{c}{Fast} \\
\cmidrule(lr){2-3}\cmidrule(lr){4-5}\cmidrule(l){6-7}
Design & L & CP & L & CP & L & CP \\
\midrule
vtr-mkPktMerge   & 0.06 & 0.64 & 0.06 & 0.80 & 0.06 & 0.85 \\
light8080        & 0.37 & 0.78 & 0.36 & 0.83 & 0.37 & 0.78 \\
epfl-adder       & 0.57 & 0.13 & 0.39 & 0.86 & 0.25 & 1.23 \\
epfl-priority    & 0.46 & 1.82 & 0.51 & 0.41 & 0.42 & 2.07 \\
epfl-voter       & 0.65 & 0.81 & 0.56 & 0.84 & 0.56 & 0.96 \\
fpu-adder        & 0.70 & 0.93 & 0.70 & 0.86 & 0.70 & 0.96 \\
t80              & 1.00 & 0.86 & 0.97 & 0.99 & 1.15 & 1.29 \\
secworks-siphash & 1.10 & 0.98 & 1.03 & 1.12 & 1.04 & 1.10 \\
itc99-b05        & 1.28 & 1.10 & 1.17 & 1.06 & 1.22 & 1.30 \\
\bottomrule
\end{tabular}
\end{table}

\paragraph*{Size}
Table~\ref{tab:qor} summarises the comparison. Of the 67 designs, 45 were built by both flows at each
goal. At every goal the open flow uses fewer logic LUTs and slices than ISE at the same goal (15--21\%
in geometric mean), and as few LUTs or fewer on 26 to 33 of the 45 designs. The ratios vary widely
between designs (Table~\ref{tab:designs}): at Area, the slices of \emph{vtr-mkPktMerge} are 0.12 of
ISE's, those of \emph{itc99-b05} 2.10 times ISE's.

\paragraph*{Speed}
At Area and Speed the open flow's critical paths are shorter than ISE's at the same goal, by 14\% and
8\%, and as short or shorter on 29 of 39 and 28 of 40 designs. At Fast, which makes one placement of the
usual mapping, they are 12\% longer, and as short or shorter on only 10 of 39. Individual designs show
why the goals matter: \emph{epfl-adder}, a 128-bit adder written as gates, has a critical path of
15.0\,ns at Area against ISE's 113.3\,ns, while \emph{epfl-priority} at Area is 1.82 times slower than
ISE and 0.41 times at Speed. Splitting each critical path that is longer than ISE's into logic and
routing, as \texttt{trce} reports it, the average excess at Area (10 designs) is 1.76\,ns of logic and
1.15\,ns of routing, and at Fast (29 designs) 2.30\,ns and 1.28\,ns; at Speed (12 designs) the logic is
on par ($-0.10$\,ns) and the excess is in routing ($+1.00$\,ns), which is where further work lies.
For the circuits of an introductory course, which are small and meet the boards' 50\,MHz clock with a
wide margin, these differences matter little.

\paragraph*{Run time}
Each flow's whole build, from synthesis to bitstream, was timed on its own machine, Silinx's in
WebAssembly, so the run times are an indication only: in geometric mean, Silinx's build took 80\% and
84\% less time than ISE's at Area and Fast, and 11\% less at Speed, which places and routes several
times.

\paragraph*{Language coverage}
XST, ISE's synthesis tool, cannot read 15 of the 67 designs (\texttt{\$clog2}, VHDL-2008 constructs such
as reading an output port, libraries other than \texttt{work}, some \texttt{for} loop forms, event
expressions in sensitivity lists); Silinx reads and builds them. Seven more designs were not built by ISE
because it rejected their pin constraints. For students, this means that HDL idioms from current
textbooks and online examples work.

\paragraph*{History}
The comparison has been run since mid-October 2026. With Silinx's earlier in-house placer and router,
the last comparison at the default effort (25 October, Silinx 15.13.0) gave 20\% fewer LUTs and 7\%
shorter critical paths than ISE.

\section{Developing with AI Coding Agents}
\label{sec:ai}

Silinx was written with Anthropic's Claude Code: the author sets the direction, files the issues, tries
the tool on the boards and reviews; Claude sessions write the code, the tests and the documentation. The
name says so: \emph{Si}, for the AI that wrote it, and \emph{linx}, for the programmable links of an
FPGA. Several sessions work in parallel, each in its own git worktree on its own issue, with one session
coordinating the merges.

The agents follow written rules, kept in the
repository and read by every session at start-up: every bug, feature or to-do is a public issue, and the
fixing change refers to it; every change ships with its own tests (unit tests compared against an
independent reference where possible, UI tests in both interfaces, generated HDL compiled and simulated
in VHDL and in Verilog, a regression test that fails on the old code); tests assert the exact warnings;
modules are kept small, so that a session reads only what it changes; the examples are the test designs;
and ISE is the oracle. Changes reach the protected main branch only through pull requests with the
continuous integration checks green, reviewed by a code owner. An external, strict oracle suits
AI-generated code: a bitstream is either byte-identical to ISE's or it is not, and a failing comparison
cannot be argued away. From its first version on 6 October 2026 to 8 November 2026, the repository's
main branch received 2\,326 commits, its issues and pull requests reached number 3\,141, and 50
releases were published.

\section{Limitations and Future Work}

The open toolchain supports the Spartan-3E boards (Basys2, Nexys2), the Spartan-6 boards (Nexys3,
Atlys) and the iCEBreaker; other boards still need ISE through the local server, and further families
are planned. The Spartan-6 flow does not yet cover clock managers' configurations beyond those measured.
Speed results are below ISE's at the Fast goal, and the remaining critical-path gap at Speed lies in
routing. The shipped device descriptions are still the ones computed from an ISE report; the measured
descriptions of the Spartan-3E dies, which avoid that dependence, pass the same checks but do not yet
replace them. The quality comparison covers the Spartan-3E only. Finally, this paper describes the
platform and its technical validation; Silinx is used in the author's courses, but a study of its use in
the classroom, and of its effect on students' learning, is future work.

\section{Conclusion}

With Silinx, Spartan-3E and Spartan-6 teaching boards stay in use, and other boards are supported too:
students design in HDL, in C or C++, or by drawing, then simulate, emulate, synthesize, place, route and
program a Basys2, Nexys2, Nexys3, Atlys or iCEBreaker from a web page, on the computer or tablet they
already have, and the Xilinx bitstreams are checked bit for bit against the vendor's own tools. Silinx is
free software (AGPL-3.0) at \url{https://github.com/pmnmalo/Silinx} and runs at
\url{https://pmnmalo.github.io/Silinx/}.

\section*{Acknowledgment}
Silinx is an independent project, not affiliated with or endorsed by AMD/Xilinx or Digilent. Xilinx,
ISE, ISim, iMPACT, Spartan and Vivado are trademarks of AMD; Basys and Nexys are trademarks of Digilent.
The code of Silinx was generated with Claude Code (Anthropic); the author directed and reviewed it and
takes responsibility for this paper.

\end{document}